\documentclass[journal=ancac3,manuscript=article]{achemso}
\setkeys{acs}{articletitle=true,etalmode=truncate,maxauthors=20}

\usepackage[version=3]{mhchem} 
\usepackage[T1]{fontenc}       
\usepackage{graphicx}
\usepackage{color}
\usepackage{float}
\usepackage[normalem]{ulem}
\usepackage{hyperref}
\usepackage{amsmath}

\newcommand{\angs}{\mbox{\normalfont\AA}}

\author{Mar\'{i}a Alfonso Moro}
\affiliation{Universit\'{e} Grenoble Alpes, CNRS, Grenoble INP, Institut NEEL, 38000 Grenoble, France}
\email{maria.alfonso-moro@neel.cnrs.fr}

\author{Yannick Dappe}
\affiliation{SPEC, CEA, CNRS, Universit\'{e} Paris-Saclay, CEA Saclay, 91191 Gif-sur-Yvette, Cedex, France}

\author{Sylvie Godey}
\affiliation{Univ. Lille, CNRS, Centrale Lille, Junia, Univ. Polytechnique Hauts-de-France, UMR 8520-IEMN, Institut d’Electronique de Micro\'{e}lectronique et de Nanotechnologie, F-59000 Lille, France}

\author{Thierry M\'{e}lin}
\affiliation{Univ. Lille, CNRS, Centrale Lille, Junia, Univ. Polytechnique Hauts-de-France, UMR 8520-IEMN, Institut d’Electronique de Micro\'{e}lectronique et de Nanotechnologie, F-59000 Lille, France}

\author{C\'{e}sar Gonz\'{a}lez}
\affiliation{Instituto de Magnetismo Aplicado (IMA-UCM-ADIF), 28230 Madrid, Spain and Departamento de F\'{i}sica de Materiales, Universidad Complutense de Madrid (UCM), 28040 Madrid, Spain}

\author{Val\'{e}rie Guisset}
\affiliation{Universit\'{e} Grenoble Alpes, CNRS, Grenoble INP, Institut NEEL, 38000 Grenoble, France}

\author{Philippe David}
\affiliation{Universit\'{e} Grenoble Alpes, CNRS, Grenoble INP, Institut NEEL, 38000 Grenoble, France}

\author{Benjamin Canals}
\affiliation{Universit\'{e} Grenoble Alpes, CNRS, Grenoble INP, Institut NEEL, 38000 Grenoble, France}

\author{Nicolas Rougemaille}
\affiliation{Universit\'{e} Grenoble Alpes, CNRS, Grenoble INP, Institut NEEL, 38000 Grenoble, France}

\author{Johann Coraux}
\affiliation{Universit\'{e} Grenoble Alpes, CNRS, Grenoble INP, Institut NEEL, 38000 Grenoble, France}
\email{johann.coraux@neel.cnrs.fr}

\title[metal-organic surface alloys]{Positional and Rotational Molecular Degrees of Freedom in a Metal-Organic Surface Alloy: the Copper-Fullerene System and its Multiple Structural Phases}

\keywords{fullerene, STM, epitaxy, DFT, energy landscape, surface alloy}

\begin{document}

%
%

\begin{abstract}
Mixing two chemical elements at the surface of a substrate is known to produce rich phase diagrams of surface alloys. Here, we extend the concept of surface alloying to the case where the two constituent elements are not both atoms, but rather one atom (copper) and one molecule (fullerene). When deposited at room temperature on a Cu(111) surface, fullerenes intermix with the metal substrate. Surprisingly, 10 distinct copper-fullerene surface alloys are found to coexist. The structure of these alloys, \textit{i.e.} their composition and commensurability relationship with the substrate, is resolved using scanning tunneling microscopy and density functional theory calculations. This diversity in the alloying process is associated to the multiple possibilities a fullerene can bind to the Cu surface. The molecules are indeed found to have in-plane and out-of-plane positional degree of freedom: the molecular alloys have elastic in-plane properties and can buckle. In addition, the molecules can rotate on their binding sites, conferring extra degrees of freedom to the system. We introduce a competing-interaction energy model, parametrized against the results of the \textit{ab initio} calculations, that describes well all the phases we observe experimentally.
\end{abstract}



\section*{Introduction}

Surface binary alloys often form at the surface of a metal when it is exposed to a different chemical element, such as another metal, a group IV atom (C, Si, Ge) or a chalcogen. \cite{Christensen,Bardi,Woodruff}. Two-dimensional (2D) alloys are involved in countless situations, either as the targeted active materials or as intermediates, sometimes parasitic phases. Relevant applications include catalysis \cite{Besenbacher,Kuld}, growth of 2D materials \cite{Lahiri,Qi,Svec,Wang}, and nano-magnetism \cite{Obrien,Honolka,Mehendale}. The physico-chemical properties of these compounds usually depend on their composition and structure, which can be ordered \cite{Mehendale,Sprunger} or disordered in the manner of a frustrated magnet \cite{Azizi,Ottaviano}.

\begin{figure}[!hbt]
 \begin{center}
 \includegraphics[width=8cm]{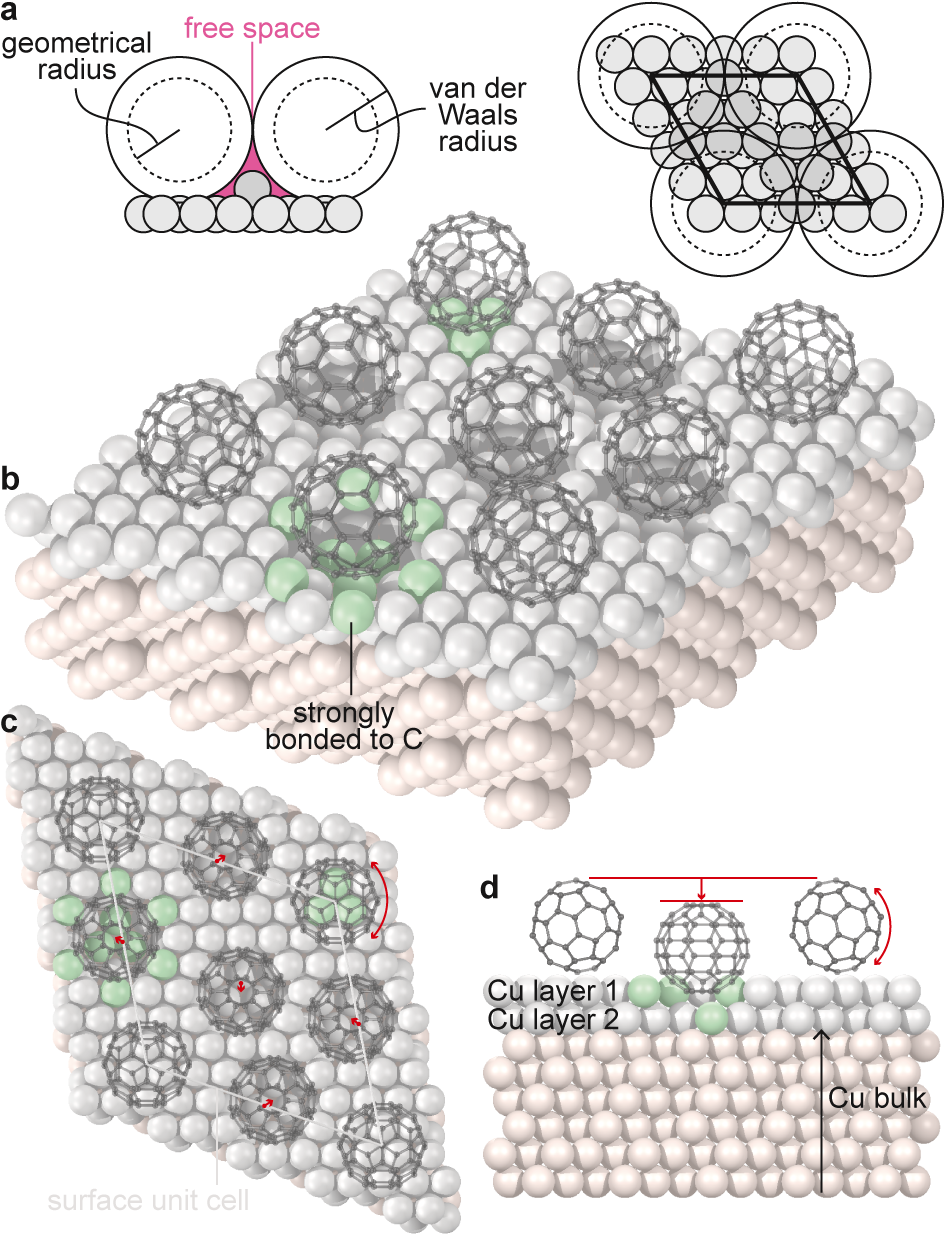}
 \caption{\label{fig0}(a) Copper-fullerene surface alloying as a ball-packing problem: atoms and molecules are represented with disks on side and top views. A free space (pink region on the side view) exists beyond the van der Waals radius of the molecules where nine Cu atoms can fit in (see top view), in the case of the commensurate supercell considered here. (b-d) Structural details of a surface alloy with a high-order commensurability on perpective (a), top (b) and cross-section (c) views. The bulk Cu atoms are represented with light orange balls, while (sub)surface Cu atoms are represented with grey balls. Green balls highlight (around two specific molecules) the Cu atoms forming stronger bonds with C atoms from the fullerenes; more bonds form in presence of Cu vacancies. The red arrows indicate the different degrees of freedom in the system.}
 \end{center}
\end{figure}

Molecules can also intermix with a metal substrate. Fullerenes (C$_{60}$) are molecules known to sink into various metal surfaces, via the formation of one-atom-deep vacancies in the substrate \cite{Stengel,Pai,Felici,Pai_b}. Interestingly, close packing of the (spherical) C$_{60}$ molecules on a flat substrate leaves an empty volume of typically 0.5~nm$^3$ (per unit cell) between the two materials. On a Cu(111) surface, this volume is readily filled by nine metal atoms (\autoref{fig0}a), in a configuration where the full Cu layer onto which the molecules sit is not the topmost one, but rather the one underneath \cite{Pai_b}. Considered this way, metal-fullerene systems are analogs of 2D alloys in biatomic systems, in what we coin \textit{metal organic surface alloys}. The two constituting elements being starkly different in size and nature, strain effects and various types of interactions (inter-molecular or with the substrate) may play a key role in the alloy stabilization at the surface while preventing bulk alloying (a common issue with biatomic alloys). Besides, the molecules may possess multiple positional and rotational degrees of freedom when lying on the metal surface (\autoref{fig0}b-d), rendering the interaction energy landcape complex.

In this work, we revisit the properties of the fullerene-Cu(111) binary system -- one of the most finely analyzed \cite{Rowe,Motai,Tsuei,Pai_c,Pai_b} hybrid metal/organic interfaces of potential relevance for future molecular electronics, and a reference system for advanced investigations of, \textit{e.g.}, the mechanics of single molecules \cite{Pawlak}, the nature of individual chemical bonds \cite{Gross} or molecular switching effects \cite{Chandler,Xu}. Specifically, we address the striking \textit{coexistence of numerous Cu-fullerene surface alloys}, whose structure is atomically resolved combining scanning tunneling microscopy (STM) measurements and density functional theory (DFT) calculations. Using DFT estimates of the different energy contributions involved in the system, and exploring the impact of the aforementioned molecular degrees of freedom, we derive an energetic classification of about 20 surface alloy candidates. Representing the surface alloy energies \textit{versus} the average inter-molecular distance and alloy composition, our model predicts that the lowest-energy phases are those observed experimentally.

\section*{Results}

\subsection*{Coexistence of surface alloys}

\begin{figure}[!hbt]
 \begin{center}
 \includegraphics[width=7.9cm]{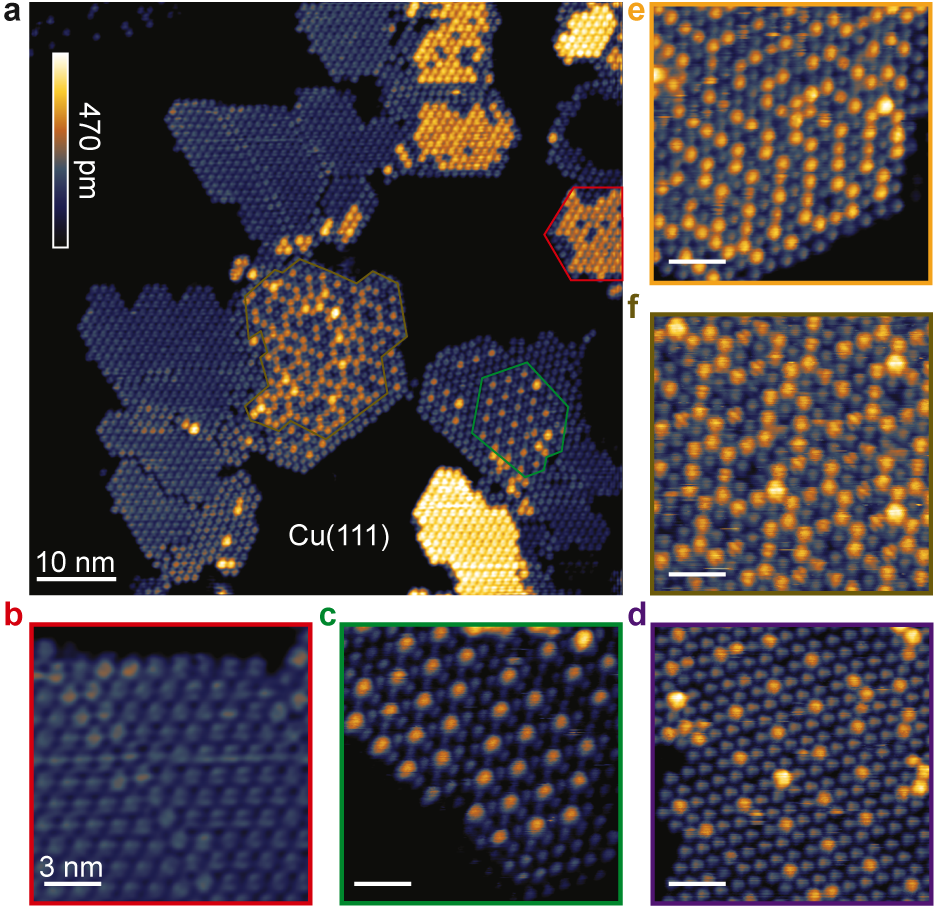}
 \caption{\label{fig1}Coexistence of 2D phases of fullerenes on Cu(111), observed with STM at room temperature in the sub-monolayer regime, (a) three phases on a single image and (b-f) zoom on five of the 10 observed phases, each framed with a specific colour. The small balls are individual molecules, the Cu(111) surface appears dark (lowest apparent height).}
 \end{center}
\end{figure}

Thermal evaporation of fullerenes under ultra-high vacuum, onto a Cu(111) surface held at room temperature (see \nameref{sec:Methods} section), produces 2D islands of densely-packed molecules. The first molecular layer grows fully before the second layer appears, presumably because several C atoms form bonds with metal atoms (\autoref{fig0}b), at the bottom pole of the molecular globe (which is not accessible to the STM tip), while fullerene diffusion onto fullerene-covered Cu(111) is a low energy-barrier-process \cite{Cuberes,Tang_b}. STM imaging reveals that 10 distinct 2D phases coexist on the surface (\autoref{fig1}a and additional data in Supporting Information Figure~S1), some of which were already observed \cite{Pai_c}. \autoref{fig1}b-f shows the five most frequent phases in our experiments, the one highlighted by a red frame (\autoref{fig1}b) being the most prominent one. Varying the fullerene flux by a factor four has no appreciable effect on the nucleation density and the relative proportion of each phase. Deposition temperature is, however, a more sensitive parameter: increasing it from $\sim$ 230~K to 300~K and 370~K (STM observations made at room temperature), we found that the phases' nucleation densities decrease, and that only the phase highlighted by a red frame in \autoref{fig1} grows at the highest temperature.

In the following, we focus on the rich energy landscape the system is able to explore at room temperature, and on the energy hierarchy between the different possible phases. To determine the energy of each phase, the precise binding configuration of each fullerene must be deciphered. This implies first to analyze the structural degrees of freedom of each molecule (see an example in \autoref{fig0}c,d), which is the purpose of the next sections. Although the positional degrees of freedom can be deduced experimentally, at least to some extent, the rotational degrees of freedom not directly accessible, since molecular orbitals, and not the carbon backbones, are imaged with STM. Finally, although geometrical considerations are used to determine the molecules' binding sites (in the presence/absence of Cu vacancies), correlation with the molecular orbitals imaged with STM (down to 4~K, see \nameref{sec:Methods} section) is required. We emphasize that the resolution of the molecular structure, which eventually relies on DFT calculations (see \nameref{sec:Methods} section), has only been done so far for the Cu-fullerene phase imaged in \autoref{fig1}b.

\subsection*{Commensurability and $xy$ positional degree of freedom}


\begin{figure}[!hbt]
 \begin{center}
 \vspace{-20pt}
 \includegraphics[width=8cm]{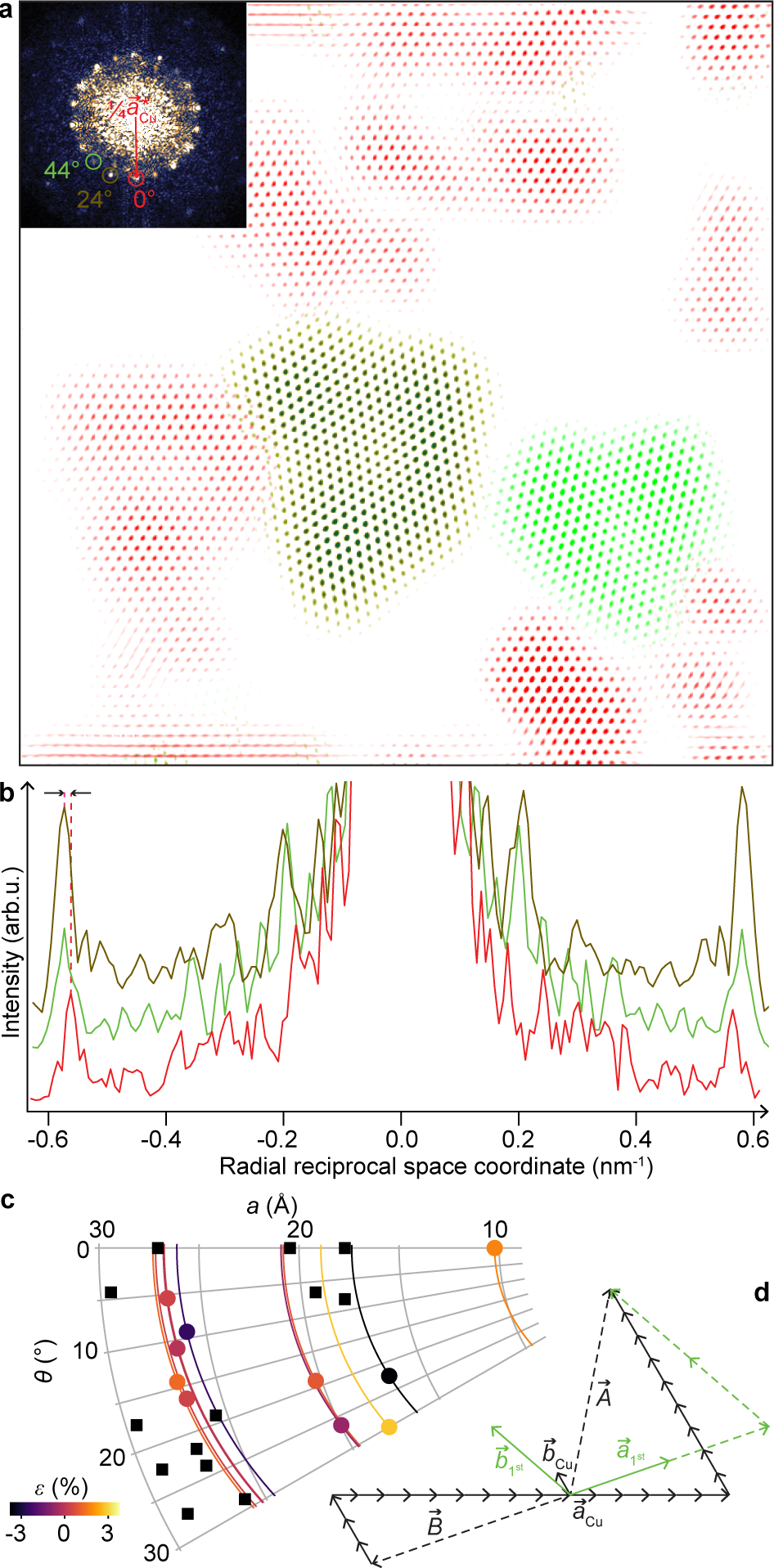}
 \caption{\label{fig2}Surface crystallography of the 2D phases. (a) Dark-field STM image overlay of \autoref{fig1}a. Inset: corresponding Fourier transform. Circles highlight one of the six (first order) harmonics used for each phase to build the dark-field image. The vector shown with a red arrow corresponds to a quarter of a substrate's reciprocal unit vector. (b) Radial profiles in reciprocal space, extracted from the inset of \textit{a} for each phase. (c) Phases' dense fullerene row orientation $\theta$ \textit{versus} lattice constant $a$ ($<$ 3.5~nm), predicted for inter-molecular strains $<$ 4\%. Some commensurate structures (disks symbols) match experiments (disk colour related to strain level); others (square symbols) are not observed in experiments. The continuous arcs are calculated for incommensurate superstructures. (d) Relationship between the unit vectors of the substrate $(\vec{a}_\mathrm{Cu},\vec{b}_\mathrm{Cu})$ and those of a commensurate phase $(\vec{A}=6\vec{a}_\mathrm{Cu}+9\vec{b}_\mathrm{Cu},\vec{B}=-9\vec{a}_\mathrm{Cu}-3\vec{b}_\mathrm{Cu})$. The green vectors connect the centers of NN fullerenes $(\vec{a}_{1^\mathrm{st}},\vec{b}_{1^\mathrm{st}})$.}
 \vspace{-20pt}
 \end{center}
\end{figure}

As mentioned above, 10 molecular phases are observed experimentally, five of which being already reported in previous works \cite{Pai_c}. Close inspection of the STM images reveals that they have distinct crystallographic orientations, i.e., each phase can be characterized by an angle $\theta$ between the dense rows of fullerenes and the dense Cu rows (see statistics over 90 fullerene-based domains, Supporting Information Section S2, Table S1). The different $\theta$ values are revealed by Fourier filtering our STM images, whereby we construct a colour-coded dark-field overlay of the phases (see an example in \autoref{fig2}a). Importantly, the first spatial harmonics do not appear at the same distance from the center of reciprocal space (compare, \textit{e.g.}, the dark green and red profiles reported in \autoref{fig2}b and derived from the Fourier transform shown in the inset of \autoref{fig2}a), \textit{i.e.} the first nearest neighbour (NN) inter-molecular distances $d_\mathrm{NN}$ are different. Note that the differences are small (a few percents at most), close to the resolution limit, and are better identified when comparing Fourier harmonics of phases that coexist on single STM images (corrected from imaging artefacts, see Section S3 in Supporting Information).

\begin{figure}[!hbt]
 \begin{center}
 \includegraphics[width=8cm]{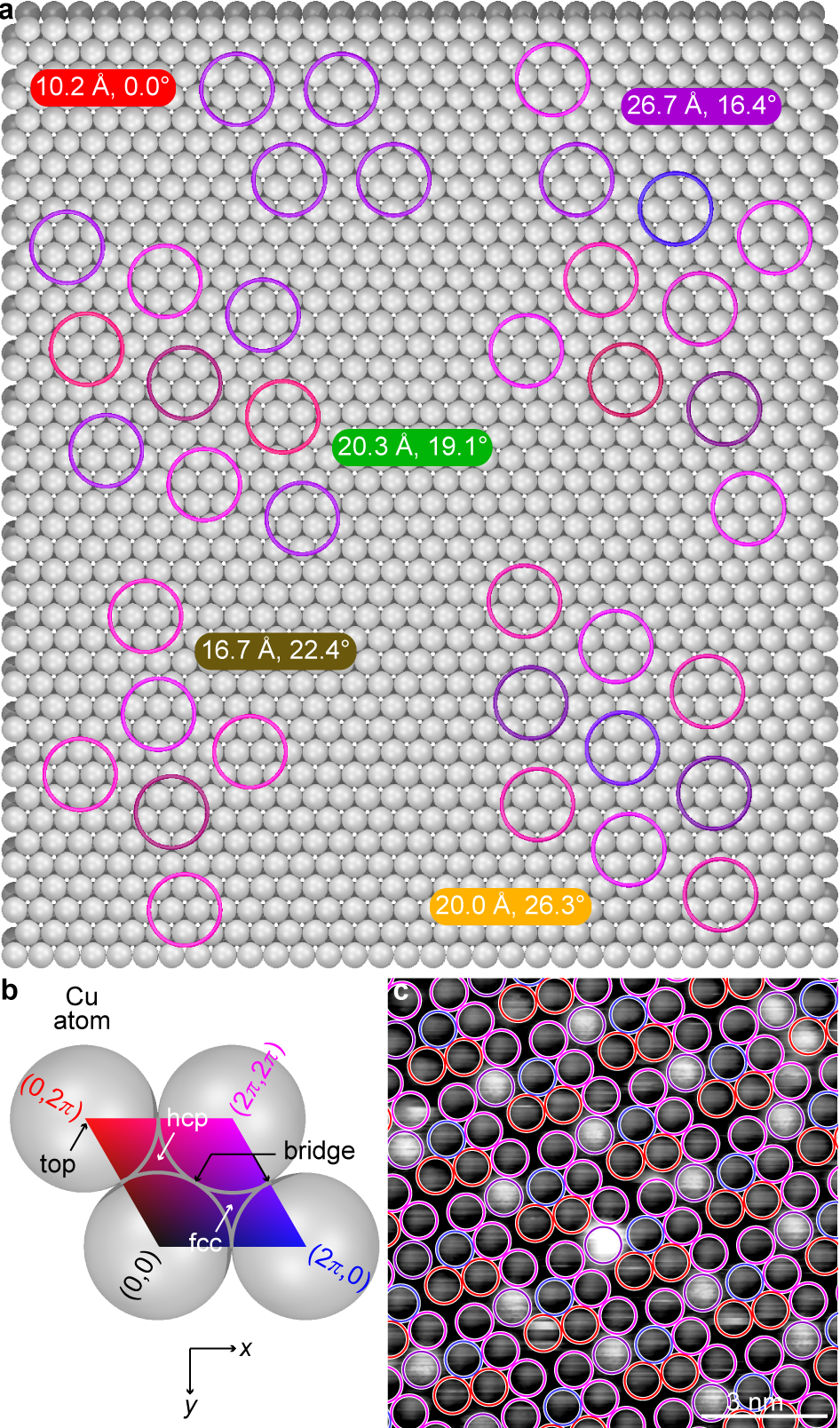}
 \caption{\label{fig3}Atomic coincidences in 2D phases and in-plane order parameter. (a) Top-view cartoons of the phases observed in \autoref{fig1}b-f, described by their lattice constant and the orientation of their dense fullerene rows. The boxes' colours match those in \autoref{fig1}b-f. There are no local variations of the NN fullerene distance within each phase. The gray disks mark the positions of Cu atoms; the larger circles are centered on the center of mass of each molecule. The circles' colour represents the molecules' geometrical phase (as defined in \textit{b}). (b) 2D map, within a Cu(111) unit cell (Cu atoms shown as gray disks), of the distance to the hcp and fcc crystallographic sites (\textit{i.e.} geometrical phase, coded with red and blue hues). (c) Local variations of the fullerene NN distance in the high-order commensurate (26.7~\AA, 16.4$^\circ$) phase, as seen with STM.}
 \end{center}
\end{figure}

From the $\theta$ and $d_\mathrm{NN}$ values of each molecular phase, a commensurate structure with the substrate can be inferred. The corresponding matrix transformation, between the unit cells of the substrate and of the phase \cite{Hooks,Artaud}, is illustrated graphically in \autoref{fig2}d, and Supporting Information Table~S2 gives the commensurability of the other observed phases. The simplest commensurate phase comprises one fullerene per unit cell, has a $\theta=0^\circ$ orientation and a lattice constant $a=10.2$~\AA\; equal to $d_\mathrm{NN}$. The latter equality does not hold for the other phases, which are so-called high-order commensurate phases \cite{Forker2017}. A real-space representation of the five phases imaged in \autoref{fig1}b-f is given in \autoref{fig3}a.

The number of conceivable commensurate structures is not limited to those observed experimentally, though, and is actually infinite in two dimensions. Limiting the remaining of the discussion to structures whose $d_\mathrm{NN}$ value remains within $\pm$4\% of the equilibrium distance (strains beyond these values would be energetically prohibitive), and leaving aside structures with unit cells beyond 30~\AA, we are still left with about 20 possible structures.

Besides, not only commensurate phases, but also a multitude of incommensurate ones may exist. Such phases would correspond to molecular lattices assuming \textit{a priori} any $\theta$ orientation (possibly stabilized by so-called static distortion waves \cite{Novaco,Meissner}). The characteristic variation expected for their lattice constant $a$ \textit{versus} $\theta$ \cite{Artaud} is shown in \autoref{fig2}c. The fact that the $\{a,\theta\}$ values of the phases we observe do not fall or distribute evenly on the incommensurate $a(\theta)$ arcs is a signature that the fullerene layer is not free-standing on its substrate. Translated in the language commonly used to describe binary compounds, the fullerene layer is not in a segregated phase onto the substrate, but rather involved in a surface alloy characterized by strong C-Cu bonds.

We now analyze the in-plane ($xy$) position degree of freedom of the molecules, \textit{i.e.} the relative position of the molecules' center and substrate atoms in the commensurate phases. For that purpose, the so-called geometrical phase, a two-component variable assuming 0-2$\pi$ variations, is employed to locate the molecules with respect to a specific crystallographic site. The geometrical phase is represented on a 2D map using a two colour channels (red and blue, see \autoref{fig3}b). If one assumes that the $d_\mathrm{NN}$ distances are all the same for a given phase, one realizes that except for the $\{10.2~\angs, 0.0^\circ\}$ commensurate phase, only part of the molecules can occupy the most favourable binding sites (whatever they are: their nature will be discussed later). This raises the question whether the molecules may move in-plane to reach more favourable sites, as proposed earlier \cite{Hsu,Pai_c}. To answer this question, we inspect STM images of the high-order $\{26.7~\angs, 16.4^\circ\}$ commensurate phase shown in \autoref{fig1}d. Converting the observed molecular positions (center of circles in \autoref{fig3}c) into a colour-coded geometrical phase, we find clear differences with the undistorted case considered in \autoref{fig3}a. This means that experimentally, the molecule position locally changes, demonstrating elastic properties \cite{Forker2017} (also found in other high-order commensurate phases). 

\subsection*{Alloying and $z$ positional degree of freedom (buckling)}

\begin{figure}[!hbt]
 \begin{center}
 \includegraphics[width=8cm]{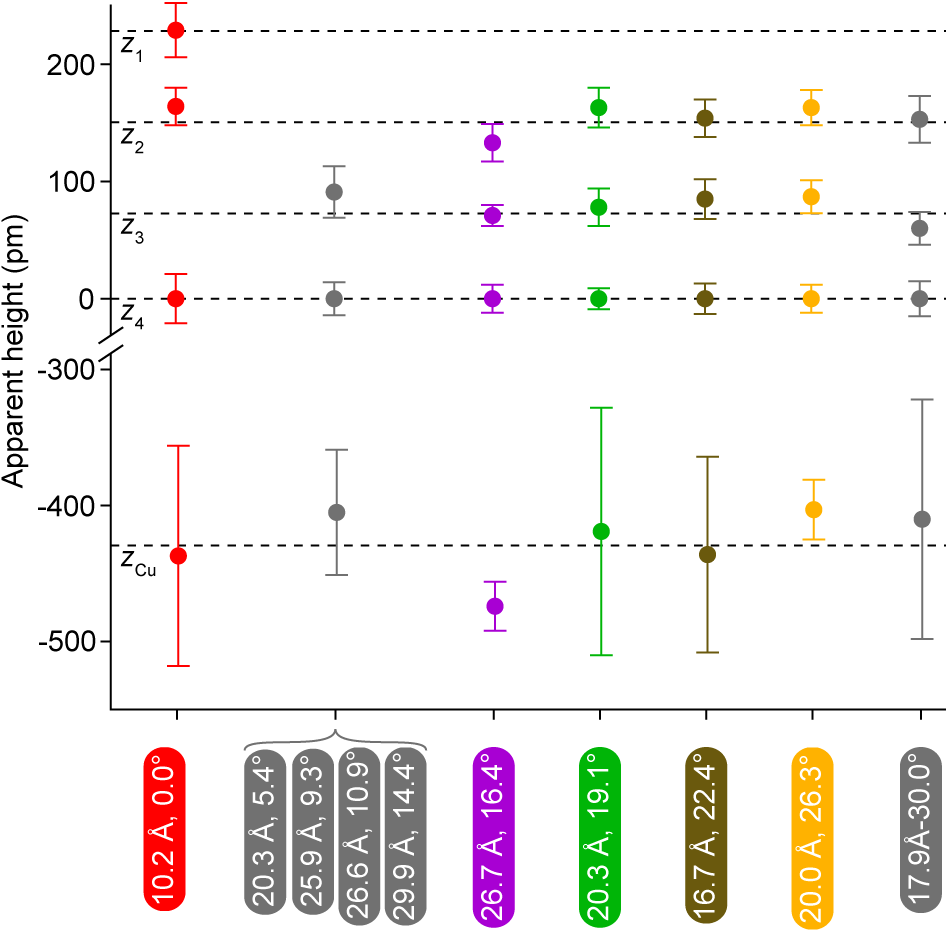}
 \caption{\label{fig4}Average apparent heights ($z_{1-4}$) of fullerene molecules in the observed phases, and of the bare Cu(111) surface ($z_\mathrm{Cu}$) around them. The bars visualize the statistical dispersion of the data.}
 \end{center}
\end{figure}

We next consider the out-of-plane ($z$) positional degree of freedom the molecules can have on the Cu(111) surface. This degree of freedom is directly apparent on the STM images as a yellow/blue contrast (see \autoref{fig1}), which demonstrates that the molecular layer buckles. Our measurements reveal that buckling can be ordered, as evidenced by the $(1\times 1)$, $(2\times 2)$ and $(\sqrt{7}\times\sqrt{7})$R19.1 patterns visible in \autoref{fig1}b-d, or disordered (\autoref{fig1}e,f). Close inspection of the STM images suggests the presence of multiple contrasts associated to different apparent heights. We emphasize that the heights assessed with STM are not true heights, since topography and local electronic density of states are entangled. To determine the height values, we thus have performed a statistical analysis on STM images measured with different tunnel bias. This parameter was, however, found to have only marginal effect on the height values. A statistical analysis of the molecules' height is summarized in \autoref{fig4} for the different phases (see also Section~S5 in Supporting Information, Table~S3).


Assigning the origin of the heights to the lower molecular level, all phases are found to exhibit three characteristic apparent height levels, of $z_4\simeq$ 0~pm, $z_3\simeq$ 80~pm, and $z_2\simeq$ 155~pm -- except for the $\{10.2~\angs, 0^\circ\}$ commensurate phase, which does not exhibit the $z_3$ level but rather a higher level at $z_1\simeq$ 220~pm. Interestingly, this phase also tends to form extended patches with uniform height pattern (\autoref{fig1}a), whereas all other phases are truly mixtures of different heights. Around all the phases, the Cu(111) surface is found at $z_\mathrm{Cu} - z_4 \simeq$ 425~pm (a large dispersion is associated to this value, due to the known unstable imaging of the Cu(111) surface at room temperature).

Overall, the $z$ position of a molecule is found to have four well-defined values $z_{1-4}$. To interpret this out-of-plane degree of freedom, two geometrical arguments can be invoked. First, the molecule's geometrical diameter is 710~pm, and the height of a Cu(111) step height is 210~pm. From these numbers, the $z_\mathrm{Cu} - z_4 \simeq$ 425~pm value corresponds to a fullerene molecule sunk into the substrate, sitting on the second Cu layer. In other words, the Cu(111) surface is embossed with one-atom-deep vacancies (each vacancy being made of seven atoms, see below) at the bottom of which the molecule sit. Second, the height difference in the simple commensurate phase (10.2~\AA, 0$^\circ$), between the low-laying molecules ($z_4$) and those found $z_4 - z_1\simeq$ 220~pm above, suggests that the latter sit on the top Cu layer and the former sit on the second Cu layer. Consistent with previous works \cite{Pai_b,Pai_c}, our conclusion is that the out-of-plane degree of freedom is intimately related to the formation of Cu vacancies. To go beyond these geometrical considerations and to complete the description of the Cu-fullerene surface alloy phases, we now need to determine the origin of the $z_{2-4}$ values.

\subsection*{Disentangling rotational and $z$ degrees of freedom with DFT}

Unambiguously determining the origin of the $z_{2-4}$ values comes down to resolve the $xy$, $z$, and rotational degrees of freedom of each molecule in the different phases. This means solving the atomic-scale structure of the system, which is not directly apparent in STM imaging. The set of possible high-symmetry configurations is unfortunately very large (64), considering
\vspace{-\topsep} 
\begin{itemize}
\setlength{\parskip}{0pt}
\setlength{\itemsep}{0pt plus 1pt}
\item the $xy$ position of the center of the molecule with respect to different possible crystallographic sites on the Cu surface, \textit{i.e.} onto a Cu atom (top), at a hollow site between three Cu atoms (two possibilities), and halfway between two Cu atoms (bridge site);
\item the presence or absence of Cu vacancies of different kind (four, see below) underneath the fullerene, hence the $z$ position;
\item the orientation of the fullerene molecule, \textit{i.e.} facing a pentagonal / hexagonal facet, an apical C atom (apex), or a C-C bond between hexagonal facets (6-6).
\end{itemize}
\vspace{-\topsep}

\noindent
To tell which of these configurations may correspond to the STM observations, knowledge on their relative energy is desirable. This information has been sought for previously with DFT calculations, performed for selected configurations (see Supporting Information, Section~S6.5), considering the $\{10.2~\angs, 0^\circ\}$ commensurate phase \cite{Wang_c,Shi_b,Xu} or small Cu clusters with individual fullerenes \cite{Larsson}. Although these earlier works highlight energy differences in terms of $xy$ and rotational degrees of freedom, and a global stabilization associated with the presence of Cu vacancies, they did not aim at a comprehensive classification of all the configurations that are relevant. We opted for a more systematic (and computationally time-consuming) approach: for reasons that will become clearer later, we selected 17 of the 64 possible configurations and calculated the binding energies with DFT calculations (Supporting Information, Section~6.1, Figure~S3, Tables S4,S5).

In the absence of Cu vacancies, the molecule with the hexagon, pentagon, apex and 6-6 orientations have lowest calculated binding energies when attached to top, top, hollow (so-called fcc), and bridge Cu sites respectively. Of all these configurations, the apex/fcc is the lowest-energy one (on this matter, see Supporting Information, Section~6.4).

Seven-atom vacancies have mostly been discussed \cite{Pai_b} in the context of the $\{10.2~\angs, 0^\circ\}$ commensurate phase prepared by mild annealing, where all molecules sit in a vacancy. Recently however, a scenario with one-atom vacancies was also considered \cite{Forcieri}. In our calculations, we considered these two kinds of vacancies, and two additional ones, where three atoms are missing (Supporting Information, Figure~S3). We considered only hexagon and pentagon molecular orientations, which are the experimentally relevant ones (see below). Except for one of the three-atom vacancies, we find substantially increased binding energy. The configuration with a pentagon on a top Cu site at the bottom of the seven-atom vacancies is about 5~eV lower in energy than the most stable site in the absence of a vacancy. The configuration with a hexagon on a top site is also very stable, slightly less though (1.5~eV). Note that creating a vacancy has a cost, related to the lower coordination of the Cu atoms expelled from the vacancy, that must be taken into account to understand the overall energy gain (Supporting Information, Section~6.2). This cost is actually only affordable, compared to other configurations without vacancies, for the seven-atom vacancies.

Another important information we derive from the DFT calculations are the true height differences between the different configurations. Without vacancies, they are of the order of tens of picometers. The molecules sink by 100 to 200~pm in the presence of the seven-atom vacancy.

\begin{figure*}[!hbt]
 \begin{center}
 \includegraphics[width=16cm]{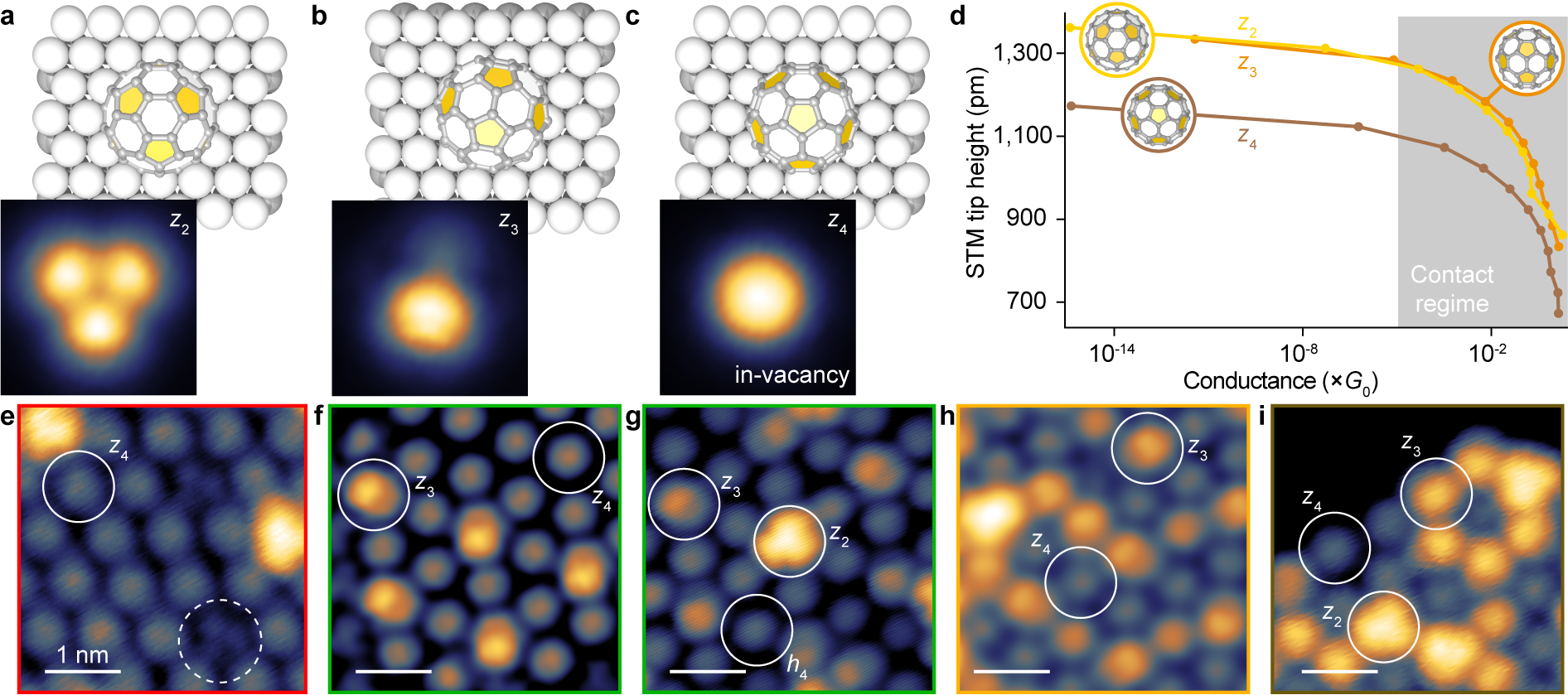}
 \caption{\label{fig5}Orientational degree of freedom of the fullerene molecules, and relation with observed apparent heights. (a-c) Atomic structure, and corresponding simulated STM images (1.5~V, 5~\AA\;tip height), for fullerene molecules in the absence of a vacancy sitting on a hexagon facet on a top site (a) and on an apex on a fcc site (b), and for a fullerene molecule sitting on a pentagon within a seven-atom Cu vacancy (c). (d) Simulated STM tip height \textit{versus} zero-bias conductance for the (a-c) configurations. (e-i) STM images measured at 4~K [1.5~V, 0.03~nA (e-g); 1.4~V, 0.01~nA (h); -1.5~V, 0.01~nA (i)], revealing the molecular orbitals of the fullerenes in four of the 2D surface alloys shown in \autoref{fig1} (coloured frames match those in \autoref{fig1}b,c,e,f).}
 \end{center}
\end{figure*}

From the above discussion we predict that the $z_2$, $z_3$, $z_4$ heights can be assigned to  hexagon/tip (no vacancy), apex/fcc (no vacancy), and pentagon/top (within a seven-atom vacancy) configurations, respectively. We simulated the STM images taking into account the effect of the tip and of the substrate (see \nameref{sec:Methods} section) for these configurations (\autoref{fig5}a-c). Distinct contrasts are clearly expected for each configuration, accounting for different spatial distributions of the electronic density of states. Calculating the conductance of the molecule-Cu tunnel junction as a function of the STM tip height for the three configurations (\autoref{fig5}d), we additionally find that beyond the regime of molecule-tip contact, at fixed conductance value the apparent height is slightly larger for the hexagon/top than for the apex/fcc configuration -- themselves substantially larger than for the in-vacancy configuration. The $\sim$130-200~pm topographic height difference in presence of the vacancy (compare Supporting Information, Tables~S4,S5) is the strongest contribution to the STM height contrast. On the contrary, the hexagon/top and apex/fcc configurations (without vacancy) differ by $\sim$70~pm in height, but not so much in apparent height, pointing to a strong contribution of electronic effects.

Experimental confirmation of the DFT predictions were obtained by high resolution low-temperature STM imaging (see \nameref{sec:Methods} section) of the different phases (\autoref{fig5}e-i), which reveals intra-molecular contrasts of different kinds (some already observed and related to the molecular orientation in specific Cu-fullerene and metal-fullerene phases \cite{Hashizume,Lu,DeMenech,Pawlak}). We find a very good match between the predictions (\autoref{fig5}a-d) and the experimental data (\autoref{fig4}, \autoref{fig5}e-i). Noteworthy we also observe that while the majority of in-vacancy molecules seem to exhibit a pentagon facet ($z_4$), a minority of them exhibits a hexagon facet (see dotted circle in \autoref{fig5}e), consistent with the lower predicted energy of the latter configuration.

We conclude, from our DFT+STM analysis, that the molecules essentially adopt two specific heights ($z_{3,4}$), occasionally a third one ($z_2$), anyhow in a one-to-one relationship with the presence ($z_4$) and absence ($z_{2,3}$) of a seven-atom, one-atom deep Cu vacancy. The molecules bind on specific crystallographic sites of the substrate with a well-defined orientation, essentially apex/fcc ($z_3$) and a pentagon/vacancy ($z_4$) configurations. In other words, the local tendency to alloying associated with the vacancy formation, \textit{i.e.} the maximization of the number of C-Cu bonds, authorizes essentially two $\left\{\mathrm{binding\;site\;(\textit{xy}) + orientational}+z\right\}$ molecular states -- these degrees of freedom being fully coupled.

\subsection*{Competing interaction model of surface alloying}

As we have seen, fullerene molecules on Cu(111) are characterized by several degrees of freedom: they can rotate and move in the three space directions to reach favourable binding sites. In addition, the Cu surface is `reactive' and has a tendency to form seven-atom vacancies that increase the number of Cu-fullerene bonds. As a consequence, a variety of structural phases (surface alloys) is expected, \textit{i.e.}, there is fairly large number of ways to establish the crystallographic orientation of the fullerene rows with respect to the Cu(111) lattice and the layer/substrate commensurability. Whether they all form or whether some are preferred is discussed below based on energy considerations.

\begin{figure*}[!hbt]
 \begin{center}
 \includegraphics[width=16cm]{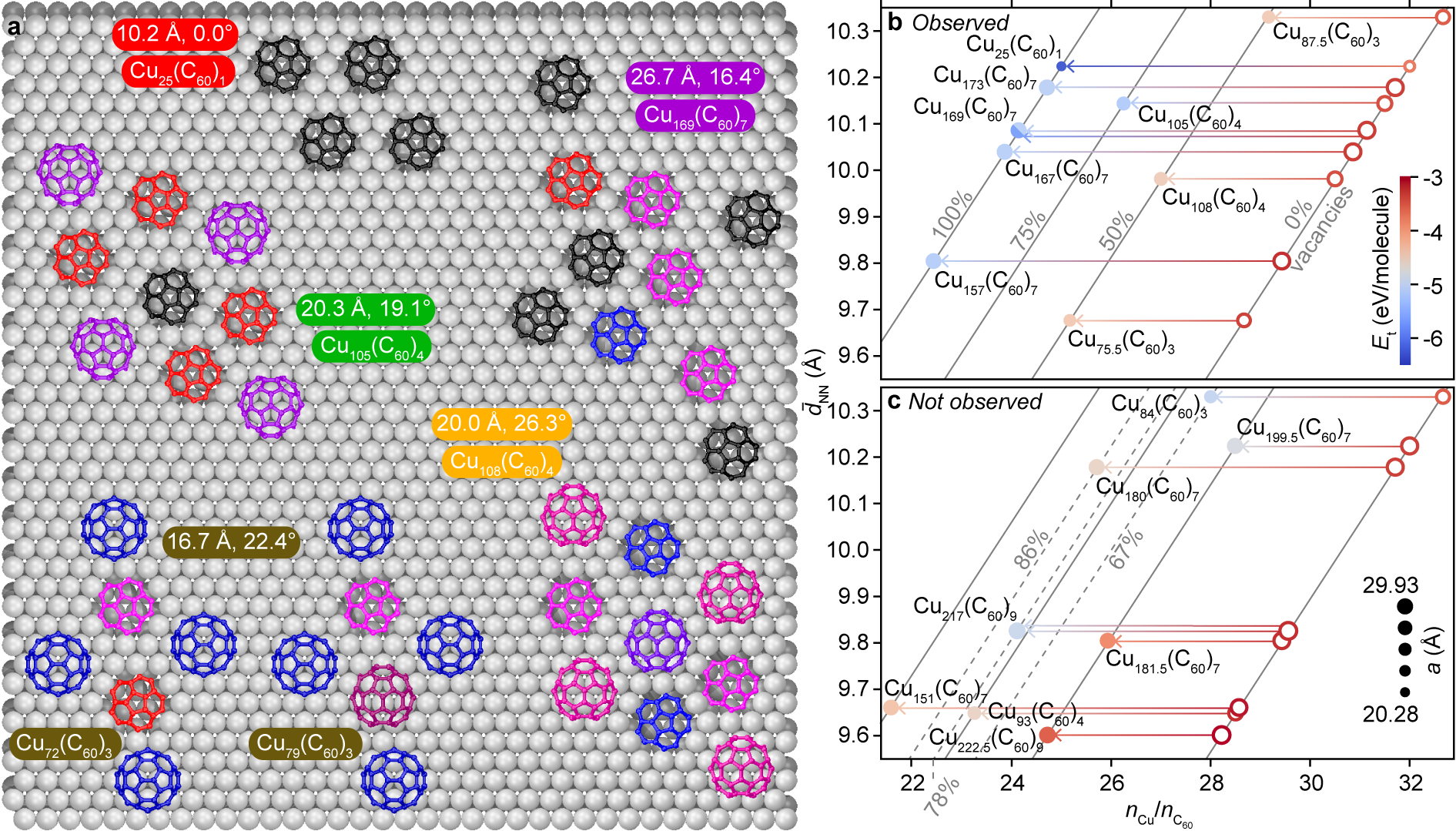}
 \caption{\label{fig6}Energetics of the 2D surface alloys. (a) Cartoons of the lowest energy configurations, obtained starting from the phases shown in \autoref{fig3}a (same colour code; red, blue, black and light pink correspond to the same crystallographic sites) and allowing 3D molecular displacements and alloying. For better clarity, only the bottom part of the molecules is shown. (b,c) Total energy ($E_\mathrm{t}$) diagram \textit{versus} composition $n_\mathrm{Cu}/n_\mathrm{C_{60}}$ and average NN fullerene distance $\bar{d}_\mathrm{NN}$, with (filled symbols) and without (open symbols) vacancies, for the observed phases (b) and those which are not observed experimentally (c). The disk size represents the surface alloys' lattice constant $a$. The colour and disk symbols are represented in the same convention/scale for \textit{b} and \textit{c}. The surface alloys have different density of vacancies, being present under 50\% to 100\% of the fullerenes.}
 \end{center}
\end{figure*}

From now on, the surface alloys will be named according to their Cu/fullerene composition. Both Cu atoms from the topmost Cu layer and Cu atoms belonging to the second (sub-surface) layer form strong bonds with fullerenes (see green-coloured Cu atoms in \autoref{fig0}b-d). The Cu content in the alloy compositions is then determined from the surface and subsurface Cu layers. Deeper Cu layers are considered to belong to the substrate (\autoref{fig0}b-d). The commensurability relationship sets the number of Cu atoms per full molecular layer, $n^\mathrm{full}_\mathrm{Cu}$; knowing the number of vacancies $n_\mathrm{vac}$ and the number of fullerenes $n_\mathrm{C_{60}}$ within the unit cell, the alloy composition is $n_\mathrm{Cu}$:$n_\mathrm{C_{60}}$ with $n_\mathrm{Cu}=2 n^\mathrm{full}_\mathrm{Cu} - 7 n_\mathrm{vac}$ (the factor 7 relates to the number of atoms per vacancy). The composition of the surface alloys observed experimentally are given in \autoref{fig6}a and in the Supporting Information, Tables~S1,S2 and Figure~S2.

To estimate the energy of each alloy phase, we now introduce a model that takes into account three, possibly antagonist, contributions: the inter-molecular pairwise interaction ($E_\mathrm{f-f}$), the dipole-dipole interaction due to an interfacial charge transfer ($E_\mathrm{d}$), and the substrate-molecule interaction ($E_\mathrm{s}$). The total energy can then be written as:

\begin{gather*}
E_\mathrm{t} (n_\mathrm{Cu},n_\mathrm{C_{60}},\bar{d}_\mathrm{NN},\theta,\{\vec{\pi}_i\}_i,\{\vec{r}_i\}_i) = \frac{1}{n_\mathrm{C_{60}}}\sum_{i\in[1,n_\mathrm{C_{60}}]} E_\mathrm{f-f}(\bar{d}_\mathrm{NN},\vec{r}_i) \\ +E_\mathrm{d}(\bar{d}_\mathrm{NN},\vec{r}_i) + E_\mathrm{s}(n_\mathrm{Cu},n_\mathrm{C_{60}},\theta,\vec{\pi}_i,\vec{r}_i)
\end{gather*}

\noindent
where the dependence on the composition, on the average NN fullerene distance $\bar{d}_\mathrm{NN}$, on the crystallographic orientation $\theta$, and on the sets of molecular orientations $\{\vec{\pi}_i\}_{i\in[1,n_\mathrm{C_{60}}]}$ ($\vec{\pi}_i$ along, \textit{e.g.}, the center-to-apex axis) and 3D displacements, $\{\vec{r}_i\}_{i\in[1,n_\mathrm{C_{60}}]}$, has been made explicit. We note that the two limit cases, in which eiher fullerene-fullerene interactions or Cu-fullerene interactions dominate, have been qualitatively addressed in a previous work \cite{Pai_c}. Here we consider all possible intermediate cases in a quantitative way, using the input of DFT calculations, and include the effect of dipole-dipole interactions.

The form of $E_\mathrm{f-f}$ is shown in the Supporting Information, Figure~S4c. It faithfully describes, with a simple analytical (hence easily parametrized) function, the result of time-dependent DFT calculations \cite{Pacheco} performed in the absence of a substrate, carefully treating the van der Waals forces dominating at large distances. $E_\mathrm{f-f}$ exhibits a marked minimum at an inter-molecular distance of 10.05~\AA, and sensitively depends, especially in the repulsive (short distance) regime, on $\{\vec{r}_i\}_i$ and $\bar{d}_\mathrm{NN}$ that vary by typically 1\% from phase to phase.

The fullerene-Cu system exhibits significant interfacial charge transfers, 1 to 3 electrons per molecule in the absence and presence of Cu vacancies respectively, according to indirect experimental estimates \cite{Pai_b}. Our DFT calculations predict a transfer of 1.2 and 1.9 electrons. The charged molecules have an image charge within the metallic substrate, implying that each molecule should be associated with an electrostatic dipole perpendicular to the surface. The distinct charge distributions, with and without vacancies, result in similar magnitude for these dipoles, 0.44~e$\times$nm and 0.48~e$\times$nm, \textit{i.e.} 0.46~e$\times$nm in average. The interaction between these dipoles ($E_\mathrm{d}$) is taken in the form of the standard electrodynamics solution assuming a one-over-cubic-inter-molecular-distance dependency. It is repulsive and its strength compares to that of $E_\mathrm{f-f}$ (Supporting Information, Figure~S4c).

The main originality of our approach, and the difficulty in the calculation of $E_\mathrm{t}$, is the evaluation of $E_\mathrm{s}$. Unlike in the common modeling of molecule-on-substrate systems, $E_\mathrm{s}$ cannot here be associated to a single form. It adopts two forms, in the presence/absence of a vacancy under the considered molecule. For that reason, we call $E_\mathrm{s}$ an \textit{adaptative} interaction energy. To parametrize it, we translate the 2D map of \autoref{fig3}b into a 2D energy landscape, by associating each high-symmetry crystallographic site to the Cu-fullerene interaction energy calculated with DFT in the presence/absence of a vacancy (Supporting Information, Figure~S4a,b).

Knowing $\vec{\pi}_i,\vec{r}_i$ for each molecule, evaluating $\sum_{i}E_\mathrm{s}$ is in principle straightforward. A challenging task however is in the determination of $n_\mathrm{vac}$, on which $\sum_{i}E_\mathrm{s}$ depends \textit{via} $n_\mathrm{Cu}$ (see above formulation of $E_\mathrm{t}$). The value of $n_\mathrm{vac}$ changes from phase to phase, not only due to different $n_\mathrm{C_{60}}$ numbers, but also due to the sometimes destabilizing interaction between neighbouring vacancies (Section~S8 in Supporting Information, Figure~S5) \cite{Shi_b}. For phases with orientation $\theta<19.1^\circ$, two neighbour vacancies cannot destabilize each other, because they are sufficiently far appart. However, the decrease of the inter-molecular distances when molecules reach their binding site within a vacancy can come with an energy penalty (see next paragraph). This penalty may be moderate, with the total energy differing by only few 10~meV for configurations with 100\% and $<$100\% vacancies, explaining the observation of both configurations in experiments (see Supporting Information, Section~S9). The energy penalty may be stronger though, preventing the formation of 100\% of vacancies for some of phases, even when $\theta<19.1^\circ$. For the (experimentally observed) phase $\theta=19.1^\circ$ orientation, we find that vacancies can only form under 75\% of the molecules without being mutually destabilized, and for the same reason, the fraction decreases to 50\% for phases with $\theta>19.1^\circ$ orientation.

To compute $E_\mathrm{t}$, we first consider all possible in-plane displacements liable to position each molecule on the nearest lowest-$E_\mathrm{s}$ configuration. This is actually what appears to happen experimentally, with molecules most often found in well-defined configurations. If the binding configuration is compatible with the formation of a vacancy, then the vacancy is formed (adaptative potential), provided that the orientation criterion introduced above is satisfied, also consistent with the STM observations. There is an important subtlety in the process of finding the proper $\{\vec{r}_i\}_i$ displacements: some of the supercells have odd multiplicity (odd $n_\mathrm{C_{60}}$), which is incompatible with half the molecules sitting at the bottom of Cu vacancies (which is the experimental observation). For such supercells we hence compute the average energies, for configurations with different numbers of vacancies per unit cell (see Supporting Information, Section~S8), and $n_\mathrm{Cu}$ is then an half-integer number.



\subsection*{Discussion: energy hierarchy between phases and coexistence}

We now compare the energies of all considered surface alloys, including those observed experimentally, when the molecular displacements and rotations are allowed and when they are not (filled/empty symbols in \autoref{fig6}b,c). Representing $E_\mathrm{t}$ \textit{versus} the composition and $d_\mathrm{NN}$, these alloys fall on parallel lines, each having a given density of Cu vacancies. In the absence of vacancies, the lower-energy phases are those with smaller unit cells. This result is actually expected from seminal epitaxial models predicting that the degree of positional coincidence with the substrate, hence the stability, should be maximized for smaller unit cells \cite{Grimmer,Smith}. However, this result does not hold anymore in the presence of vacancies as the energy strongly decreases. The reason is a strengthened binding ($E_\mathrm{s}$ gain) associated to the vacancies, causing changes in the inter-molecular distances, thus of $E_\mathrm{f-f}+E_\mathrm{d}$. The energy span between the different phases is substantially extended in presence of the vacancies, reaching almost 2~eV. Strikingly, for each experimentally relevant vacancy density, according to our model the observed alloys have lower energies than other possible alloys (compare \autoref{fig6}b and \autoref{fig6}c for vacancy densities of 100\%, 75\% and 50\%; the only exception being the not-observed Cu$_{199.5}$(C$_{60}$)$_7$ phase). In that sense the agreement with the experiments is good.

Interesting, the 169:7 and 217:9 alloy compositions both correspond to two possible phases (5.4$^\circ$ / 16.4$^\circ$ and 4.3$^\circ$ / 17.5$^\circ$ orientations respectively) with different predicted $E_\mathrm{t}$ (-5.76~eV / -5.14~eV and -4.97~eV / -4.93~eV; values expressed per molecule). This shows that, unlike in biatomic surface alloys, the composition alone is not sufficient to understand the alloy energetics. However, the crystallographic orientation of the alloy seems a relevant parameter as $E_\mathrm{t}$ increases with it for the same composition.

The phases that are not observed experimentally have vacancy densities lower than 100\% and are usually not, given the fraction of molecules within a vacancy, the lowest-energy configurations. Besides, three of the non-observed phases (Cu$_{217}$(C$_{60}$)$_9$, $\times$2, and Cu$_{222.5}$(C$_{60}$)$_9$) have higher-order commensurability supercells, with large lattice constant (symbol size codes the lattice constants in \autoref{fig6}b,c) close to 30~\AA. Forming such phases would imply a concerted assembling of a large number of fullerenes and vacancies, which is statistically unlikely, explaining the absence of the Cu$_{217}$(C$_{60}$)$_9$ surface alloys in the measurements despite their relatively low energy. The same argument can explain why the Cu$_{199.5}$(C$_{60}$)$_7$ phase, whose energy is smaller by $\sim$0.3~eV/molecule than that of the Cu$_{75.5}$(C$_{60}$)$_3$, Cu$_{108}$(C$_{60}$)$_4$ and Cu$_{87.5}$(C$_{60}$)$_3$ phases having the same vacancy density, is absent in experiments. We note that the Cu$_{84}$(C$_{60}$)$_3$ alloy could not be found in our STM images, although it has a relatively low energy and a rather small unit cell. It has a vacancy density of 2/3, and for similar orientations, lower energy phases are predicted by our model (and in fact observed experimentally, \textit{e.g.} Cu$_{157}$(C$_{60}$)$_7$). Whether accumulating yet more microscopy data would be necessary to observe this phase remains open questions.

We now address the question of phase coexistence. We first discuss the model's approximations made to treat the problem with a reasonably small set of parameters (evaluated with DFT rather than freely adjusted). One important effect of the model is the formation of Cu vacancies. As explained in the previous section, we relate $n_\mathrm{vac}$ to the orientation $\theta$ of the dense fullerene rows. We came up with a criterion on $\theta$, but did not consider the energy cost or gain. Intuitively, this cost/gain should depend on the relative position of the vacancies, since, for instance, Cu atoms can belong to the edges of two vacancies (see Section~S7 in Supporting Information, Figure~S4) \cite{Shi_b}. Disentangling the corresponding energy contribution from others is challenging in DFT calculations. In addition, such a DFT analysis would require to compare calculations made with large supercells (\textit{e.g.}, for the Cu$_{169}$(C$_{60}$)$_7$ phase with $\theta=16.4^\circ$ and $a=26.7$~\AA), and all having different $n_\mathrm{Cu}$ and $n_\mathrm{vac}$ numbers, overall making the comparison almost unfeasible. This is why in our periodic boundary conditions DFT calculations, individual vacancies (together with one fullerene) were considered within a relatively large supercell, limiting the interactions between neighbouring supercells. In the actual phases, the vacancies+fullerenes may cross-talk through the substrate (and no only by van der Waals or dipole-dipole interactions), which we do not take into account. The energy of a C-Cu bonds formed within vacancies could then differ from our estimate, and so could the energy span between the different phases.

Yet, the energies of some of the phases will still be different, by few to several 100~meV. Such large differences prompt why a phase coexistence is observed experimentally. As discussed in the Supporting Information, Section~S10, the Gibbs energy of the system comprise an entropy term, which represents here a small contribution that can only marginally modify the relative energies of the different phases. With this in mind, we can propose a growth-by-phase-coexistence scenario. We remind that the phase coexistence is observed at room temperature and slightly above it. A growth temperature increase of about 10\% leads to the exclusive formation of the Cu$_{25}$(C$_{60}$)$_1$ phase. The entropy term $-TS$ in the Gibbs free energy increases by 10\% accordingly, hence a limited increase of the entropy stabilisation is expected. In contrast, temperature can influence the dynamics of domain boundaries (\textit{i.e.}, defects) between phases and/or the kinetics of the interconversion of the different phases into the most stable Cu$_{25}$(C$_{60}$)$_1$ phase. The probability to cross the corresponding energy barrier, according to an Arrhenius law, may increase by much more than 10\%, depending on the value of the energy barrier, which would explain both the full conversion of the different phases, whose total energies span over about 2~eV, into the Cu$_{25}$(C$_{60}$)$_1$ phase at 370~K, and the coexistence at 300~K.

\section*{Summary and prospects}

We showed that the fullerene-copper system forms what we coined metal-organic surface alloys, reminiscent of more commonly known biatomic surface alloys. The variety of alloys observed experimentally was found to be associated with the multiple degrees of freedom each fullerene molecule can have on the metal surface: in-plane and out-of-plane positional displacements, as well as internal rotation. We then resolved the binding configurations of the molecules based on high-resolution scanning probe microscopy and extensive first principles calculations. The alloying mechanism is systematically accompanied with the formation of one-atom-deep vacancies in the copper surface. To interpret the coexistence of metal-organic surface alloys, we developed a model that accounts for the inter-molecular interactions and the binding to the substrate in the presence or absence of Cu vacancies. Having considered (marginal) entropy contributions to the Gibbs free energy of the system, we finally rationalized the observed coexistence of phases.

The methodology we developed is relevant to explore various metal-organic systems, including other metal-fullerene systems \cite{Altman_b,Motai,Altman,Wang_d,Sakurai_a,Sakurai_b,Johansson,Maxwell,Maxwell_b,Rogero,Guo,Stengel,Zhang_b,Schull,Li,Tang,Gardener,Pussi,Passens}. Our approach sheds light on the heteroepitaxy process involving a hybrid molecular/inorganic system on either sides of the interface. Beyond the vacancy-induced alloying mechanism, which may be seen as the premises of intermixing, the nature of heteroepitaxy could be further explored by considering the elastic properties of the fullerene lattice, for instance seeking for possible static distorsion waves \cite{Novaco} that were observed in non-alloyed metal-organic interfaces \cite{Meissner}. Finally, some of the phases with non-trivial disorder show striking similarities with frustrated 2D spin systems. In this sense the Cu-fullerene system is also a rich playground to investigate the many-body physics often associated with frustrated magnetism, and more specifically the so-called triangular Ising antiferromagnet \cite{Wannier}.

\section*{Methods}
\label{sec:Methods}

\subsection*{Sample preparation}

Experiments were performed under ultra-high vacuum, inside two chambers (one in Grenoble, the other in Lille), each with base pressures in the low 10$^{-10}$~mbar range. Copper single-crystals, cut and polished along a (111) surface, bought from Surface Preparation Laboratory, were prepared by repeated cycles of Ar$^{+}$ ion sputtering (0.8-1.0~keV) and moderate temperature annealing (620-700~K). The surface quality was checked with electron diffraction and STM. Fullerene molecules (99.9\% purity) were purchased from Sigma Aldrich. They were thoroughly out-gassed under ultra-high vacuum, within a Kentax evaporator. Evaporation was performed on clean Cu(111) at room temperature and 173-223~K, at rates of 0.5-2.0~\AA/min using the water-cooled evaporator held at a temperature of around 620-640~K.

\subsection*{Scanning tunneling microscopy}

Scanning tunneling microscopy was performed with two different instruments, (one in Grenoble, the other in Lille), both connected under ultra-high vacuum to the growth chambers. For room temperature measurements a STM-1 Omicron microscope was used, with electrochemically-etched W metal probes. For low temperature (4~K) measurements, experiments were carried out with a Joule-Thomson microscope (SPECS, Berlin) using a commercial length-extensional resonator (Kolibri, SPECS Berlin) with W tip prepared by Ar$^+$ sputtering and substrate conditioning.

Correction of drift and hysteresis imaging artefacts are discussed in Section~S3 of the Supporting Information.

\subsection*{Density functional theory calculations}

The structures and binding energies of the fullerene molecules on their Cu(111) substrate were optimized using the localized-orbital density functional theory Fireball code \cite{Sankey,Lewis,Jelinek}, within the local density approximation. In our periodic boundary conditions simulations, the substrate was modeled with a five-layer stack of Cu atoms, imposing a fixed position for atoms in the bottom two layers to simulate bulk layers. Optimized minimal basis sets of localized orbitals, $sp^3$ for carbon atoms and $sp^3d^5$ for copper atoms, were used for structure optimization, with respective cutoff radii (in atomic units) 4.5 ($s$), 5.7 ($p$), 3.7 ($d$), for Cu, and 4.5 ($s$), 4.5 ($p$), for C, following a previous work on fullerene/Cu(111) \cite{Schull_b}. Also, an $sp^3d^5$ basis set of carbon orbitals has been used for a better description of charge transfer and electronic dipoles. Different sizes of supercells have been considered in this work to simulate different inter-molecular distances. All the structures have been fully optimized, using a set of 8$\times$8$\times$1 $k$-points in the Brillouin zone until forces remained below 0.1~eV/\AA.

The tunneling current in the STM simulations was calculated using the non-equilibrium Green function formalism developed by Keldysh \cite{Keldysh,Mingo}. First, we optimized the structure and determined the electronic properties of a Cu-tip formed by a pyramid of four atoms joined to four layers of Cu(111) using the same DFT calculation scheme as for the different C$_{60}$/Cu(111) configurations. In a second step, the tip was placed above the upmost C atom of the fullerenes, at a distance of 5~\AA\; from it, for different fullerene/Cu(111) configurations optimized with the DFT simulations and avoiding multiple scattering effects. The images presented in \autoref{fig5}f-h were calculated with a voltage or +1.5~V, corresponding to the empty states of the molecule. A complete description of the methodology can be found elsewhere \cite{Gonzalez}.

\begin{suppinfo}
The Supporting Information is available free of charge at \href{https://pubs.acs.org.doi/XXXXX}{https://pubs.acs.org.doi/XXXXX}.
\begin{itemize}
\item Supplementary figure (Supplementary Figure~1) on additional STM data of the different phases
\item Supplementary table (Supplementary Table~1) on the statistical analysis of the orientation of the phases
\item Supplementary table (Supplementary Table~2) on the nomenclatures that can be used to describe the different phases
\item Supplementary figure (Supplementary Figure~2) showing cartoon views of the 10 observed phases
\item Supplementary table (Supplementary Table~3) on the statistical analysis of the molecules' height as measured with STM
\item Supplementary figure (Supplementary Figure~3) on DFT calculations of the energy \textit{versus} Cu-molecule distance for various binding configurations
\item Supplementary tables (Supplementary Tables~4-6) summarizing the binding energies, molecular heights for various molecule-Cu binding configurations
\item Supplementary figure (Supplementary Figure~4) showing the 2D molecule-Cu interaction potential with and without vacancies and the inter-molecular interaction potential together with the dipole-dipole interaction \textit{versus} inter-molecular distance
\item Supplementary figure (Supplementary Figure~5) showing cartoons of the possible arrangement of Cu vacancies in different phases observed experimentally
\item Supplementary figure (Supplementary Figure~6) giving a histogram of the density of vacancies per unit cell in the Cu$_{75.5}$(C$_{60}$)$_3$ surface alloy
\item Supplementary figure (Supplementary Figure~7) representing two possible configurations for a given phase
\item Discussions about each of the Supplementary Figures and Tables, and about the correction of STM imaging artefacts, on different possible kinds of Cu vacancies, on the effect of the supercell size in the DFT calculations, on the evaluation of the energy cost for vacancy formation, on previous DFT work on Cu-fullerenes, on the possible interactions between neighbour vacancies, on the way disorder can be taken into account in our competing interaction model, and on the coexistence of two structures for spectific phases
\end{itemize}

\end{suppinfo}

\section{Author Information}

\subsection*{Corresponding Authors}

\textbf{Mar\'{i}a Alfonso Moro} -- \textit{Universit\'{e} Grenoble Alpes, CNRS, Grenoble INP, Institut NEEL, 38000 Grenoble, France},\\
\href{http://orcid.org/0000-0003-3906-4899}{http://orcid.org/0000-0003-3906-4899};\\
Email: \href{mailto:maria.alfonso-moro@neel.cnrs.fr}{maria.alfonso-moro@neel.cnrs.fr}

\noindent
\textbf{Johann Coraux} -- \textit{Universit\'{e} Grenoble Alpes, CNRS, Grenoble INP, Institut NEEL, 38000 Grenoble, France},\\
\href{http://orcid.org/0000-0003-2373-3453}{http://orcid.org/0000-0003-2373-3453};\\
Email: \href{mailto:johann.coraux@neel.cnrs.fr}{johann.coraux@neel.cnrs.fr}

\subsection*{Authors}

\textbf{Yannick Dappe} -- \textit{SPEC, CEA, CNRS, Universit\'{e} Paris-Saclay, CEA Saclay, 91191 Gif-sur-Yvette, Cedex, France}

\noindent
\textbf{Sylvie Godey} -- \textit{Univ. Lille, CNRS, Centrale Lille, Junia, Univ. Polytechnique Hauts-de-France, UMR 8520-IEMN, Institut d’Electronique de Micro\'{e}lectronique et de Nanotechnologie, F-59000 Lille, France}

\noindent
\textbf{Thierry M\'{e}lin} -- \textit{Univ. Lille, CNRS, Centrale Lille, Junia, Univ. Polytechnique Hauts-de-France, UMR 8520-IEMN, Institut d’Electronique de Micro\'{e}lectronique et de Nanotechnologie, F-59000 Lille, France}

\noindent
\textbf{C\'{e}sar Gonz\'{a}lez} -- \textit{Instituto de Magnetismo Aplicado (IMA-UCM-ADIF), 28230 Madrid, Spain and Departamento de F\'{i}sica de Materiales, Universidad Complutense de Madrid (UCM), 28040 Madrid, Spain}

\noindent
\textbf{Val\'{e}rie Guisset} -- \textit{Universit\'{e} Grenoble Alpes, CNRS, Grenoble INP, Institut NEEL, 38000 Grenoble, France}

\noindent
\textbf{Philippe David} -- \textit{Universit\'{e} Grenoble Alpes, CNRS, Grenoble INP, Institut NEEL, 38000 Grenoble, France}

\noindent
\textbf{Benjamin Canals} -- \textit{Universit\'{e} Grenoble Alpes, CNRS, Grenoble INP, Institut NEEL, 38000 Grenoble, France}

\noindent
\textbf{Nicolas Rougemaille} -- \textit{Universit\'{e} Grenoble Alpes, CNRS, Grenoble INP, Institut NEEL, 38000 Grenoble, France}

\subsection*{Author contributions}

J.C. and N.R. conceived the projet. M.A.M and S.G. performed the experiments with support from V.G.,  P.D., T.M. and J.C. Y.D. performed the DFT calculations and C.G. performed the STM image simulations. M.A.M and J.C analyzed the results. All authors wrote and revised the paper.

\subsection*{Note}

The authors declare no competing financial interest.

\begin{acknowledgement}

M.A.M. thanks the Nanosciences Foundation, hosted by the Universit\'{e} Grenoble Alpes Foundation, for a Ph.D. grant, and Groupement de Recherche NS-CPU (Nanosciences en champ proche sous ultravide) for funding of accommodation fees during measurement campaigns in Lille. This work was carried out in part using the SPM facilities of the Multi-Physics Characterisation Platform (PCMP) of the IEMN and the RENATECH network. We thank Dr. Alexandre Artaud for assistance in the correction of imaging artefacts in STM images. CG acknowledges the computer resources at Lusitania II and the technical support provided by the Centro Extreme\~{n}o de iNvestigaci\'{o}n, Innovaci\'{o}n Tecnol\'{o}gica y Supercomputaci\'{o}n (CENITS), project FI-2022-1-0020 and funding by Comunidad de Madrid (Spain), project S2018/NMT-4321 NANOMAGCOST. 

\end{acknowledgement}

\newpage


\providecommand{\latin}[1]{#1}
\makeatletter
\providecommand{\doi}
  {\begingroup\let\do\@makeother\dospecials
  \catcode`\{=1 \catcode`\}=2 \doi@aux}
\providecommand{\doi@aux}[1]{\endgroup\texttt{#1}}
\makeatother
\providecommand*\mcitethebibliography{\thebibliography}
\csname @ifundefined\endcsname{endmcitethebibliography}
  {\let\endmcitethebibliography\endthebibliography}{}

\end{document}